\documentclass[hidelinks, 10pt, conference, letterpaper]{IEEEtran}
\usepackage{setspace}
\usepackage{graphicx}
\usepackage{array}
\usepackage{xcolor}
\usepackage{color, colortbl}
\usepackage{epstopdf}
\usepackage{enumitem}
\usepackage{multirow}

\makeatletter
\def\set@curr@file#1{%
  \begingroup
    \escapechar\m@ne
    \xdef\@curr@file{\expandafter\string\csname #1\endcsname}%
  \endgroup
}
\def\quote@name#1{"\quote@@name#1\@gobble""}
\def\quote@@name#1"{#1\quote@@name}
\def\unquote@name#1{\quote@@name#1\@gobble"}
\makeatother

\usepackage[hyphenbreaks]{breakurl}
\usepackage[hyphens]{url}

\usepackage{amsfonts}
\usepackage{dsfont}

\usepackage{mathtools}

\usepackage{amsthm}

\usepackage{graphicx}

\usepackage{epstopdf}

\usepackage{subfig}
\usepackage{caption}

\usepackage{float}

\usepackage{epsfig}

\usepackage{array,longtable}
\renewcommand*{\arraystretch}{1.0}

\usepackage{mdframed}

\newtheorem{definition}{Definition}[]

\newtheorem{theorem}{Theorem}[]

\usepackage{multirow}

\usepackage{url}

\PassOptionsToPackage{hyphens}{url}\usepackage{hyperref}

\usepackage{epsfig}

\usepackage{array,longtable}
\renewcommand*{\arraystretch}{1.0}

\newcommand{\kww}[1]{{\color{blue}\texttt#1}}

\newcommand{\holl}{\textsf{HOL Light}}
\newcommand{\mlab}{\textsf{MATLAB}}
\newcommand{\fasm}{\textsf{FASiM}}
\newcommand{\ml}{\textsf{ML}}
\newcommand{\xml}{\textsf{XML}}
\usepackage{hyperref}
\usepackage{mdframed}

\usepackage{amsmath}
\usepackage{amssymb}
\usepackage{float}
\usepackage[belowskip=-5pt,aboveskip=15pt]{caption}
\definecolor{LightCyan}{rgb}{0.88,1,1}
\makeatletter

\newcommand*{\@rowstyle}{}

\newcommand*{\rowstyle}[1]{
  \gdef\@rowstyle{#1}%
  \@rowstyle\ignorespaces%
}

\newcolumntype{=}{
  >{\gdef\@rowstyle{}}%
}

\newcolumntype{+}{
  >{\@rowstyle}%
}
\makeatother
\begin{document}
\bstctlcite{IEEEexample:BSTcontrol}
\pagenumbering{gobble}
\title{\fasm: A Framework for Automatic Formal Analysis of Simulink Models of Linear Analog Circuits}

\author{\IEEEauthorblockN{Adnan Rashid, Ayesha Gauhar and Osman Hasan
}
\IEEEauthorblockA{School of Electrical Engineering and Computer Science (SEECS) \\
National University of Sciences and Technology (NUST), Islamabad, Pakistan\\
Email: \{adnan.rashid, 14mseeagauhar, osman.hasan\}@seecs.nust.edu.pk}
\vspace{-1.3cm}
}

\maketitle
\begin{spacing}{0.98}
\end{spacing}
\addtolength{\oddsidemargin}{-.1in}
\addtolength{\evensidemargin}{-.1in}
\addtolength{\textwidth}{0.22in}
\addtolength{\topmargin}{-.1in}
\addtolength{\textheight}{0.22in}
\begin{abstract}
Simulink is a graphical environment that is widely adapted for the modeling and the Laplace transform based analysis of linear analog circuits used in signal processing architectures. However, due to the involvement of the numerical algorithms of \mlab~in the analysis process, the analysis results cannot be termed as complete and accurate. Higher-order-logic theorem proving is a formal verification method that has been recently proposed to overcome these limitations for the modeling and the Laplace transform based analysis of linear analog circuits. However, the formal modeling of a system is not a straightforward task due to the lack of formal methods background for engineers working in the industry. Moreover, due to the undecidable nature of higher-order logic, the analysis generally requires a significant amount of user guidance in the manual proof process. In order to facilitate industrial engineers to formally analyze the linear analog circuits based on the Laplace transform, we propose a framework, \fasm, which allows automatically conducting the formal analysis of the Simulink models of linear analog circuits using the \holl~theorem prover. For illustration, we use \fasm~to formally analyze Simulink models of some commonly used linear analog filters, such as Sallen-key filters.
\end{abstract}




\section{Introduction}\label{SEC:Intro}

In the past couple of decades, the analog and mixed signal and integrated circuit design has experienced dramatic changes due to their wider utility in the embedded systems, such as mobile phones, cameras and other electronic appliances based on the growing demands of the consumer marketplace. These changes involve reduction in their sizes, operating voltages and power consumption capabilities, which resulted in to an enormous increase in complexities of these circuits due to the placement of a large amount of analog and digital components on a small area. Due to this rapidly growing market of the embedded systems and complexities of their corresponding design, the accuracy of the design and the associated analysis has become a dire need, since a small design flaw can result into undesirable consequences, like heavy financial losses that can directly effect the company developing these designs and the market as well.

\begin{figure}[!ht]
\centering
\scalebox{0.442}
{\hspace*{-0.31cm}\includegraphics[trim={5.0 0.3cm 5.0 0.3cm},clip]{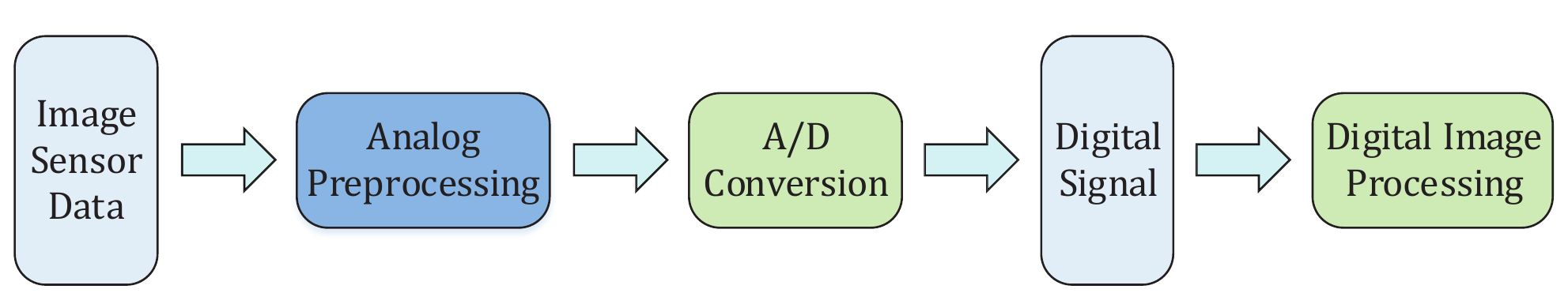}}
\caption{Image Sensor based Signal Processing Architecture}
\label{FIG:sp_arch}
\end{figure}

Linear analog circuits are a vital part of the integrated circuits and their accurate design and analysis is of utmost importance. For example, they are a fundamental component of a Complementary Metal–oxide–semiconductor (CMOS) image sensor based digital camera, which provides significant advantages in comparison to the traditional Charge-coupled Device (CCD) imaging systems~\cite{shi2015concept,soell2015cmos}. Figure~\ref{FIG:sp_arch} presents a typical image sensor based signal processing architecture that is used for imaging and mixed-signal processing. It involves an analog signal acquisition from the image sensor. To improve the quality of the image by reducing the random image noise, some analog preprocessing algorithms, composed of linear analog circuits (spatial filters), are applied. The resultant analog signal is converted to a digital signal using an A/D converter, which is then used for further digital image processing~\cite{shi2015concept,soell2015cmos}. Being a major part of a image processing architecture, there is a dire need of accurate design and analysis of linear analog circuits.

One of the most challenging aspects of modeling and analyzing linear analog circuits is to capture their continuous dynamics in their corresponding mathematical models, which are based on physical laws. For example, in order to model the voltages and currents passing through the electrical components and their interaction, we have to use circuit's governing laws, such as Kirchhoff’s Voltage Law (KVL) and Kirchhoff’s Current Law (KCL). These mathematical models are further used to derive the corresponding linear differential equations of the underlying system. The next step is to solve these equations to study properties of the system, such as transfer function and frequency response. However, solving these equations in time domain is not quite straightforward due to the involvement of the differential and integral operators. The Laplace transform~\cite{decarlo1995linear}, an integral based transform method, is often used to solve these equations by converting the time domain models (differential equations) to their corresponding algebraic equations in the frequency domain, i.e., the $s$-domain, which allows us to transform the integral and differential operators to their corresponding division and multiplication operators. These equations can now be solved in a quite straightforward way to obtain the corresponding transfer function~\cite{decarlo1995linear} and frequency response~\cite{decarlo1995linear} of the system or solution of the differential equations in frequency domain. Finally, the uniqueness of the Laplace transform is utilized to obtain the solution of these differential equations in time domain. All these properties play a crucial role in designing a reliable analog circuit.

Simulink~\cite{matlab18simulink} is a graphical environment that provides the modeling of a system in the forms of blocks and supports its corresponding analysis using \mlab. It has been widely used for modeling of analog circuits and performing their transfer function based analysis~\cite{franco1995electric}. To perform the Laplace transform based analysis of a linear analog circuit, we generally develop its Simulink model, which is further analyzed to study its various characteristics, such as transfer function and frequency response. Moreover, the theory of the Laplace transform, available in \mlab, can be used for analyzing the continuous dynamics of the circuits. For example, the \mlab~function \texttt{\textsf{laplace (f)}} accepts a time domain function \texttt{\textsf{f}} or any linear differential equation and provides its corresponding algebraic expression, which can further simplified to obtain the corresponding transfer function, i.e., the system's input-output relationship in the $s$-domain. This transformation of the time domain model to its corresponding algebraic expressions in $s$-domain can also be obtained using some other variants of \texttt{\textsf{laplace}}, i.e., \texttt{\textsf{laplace (f,transVar)}} and \texttt{\textsf{laplace (f,var,transVar)}} that provide the flexibility for the selection of the time domain variable and corresponding frequency domain variable other than $t$ and $s$, respectively. Similarly, the function \texttt{\textsf{ilaplace (F)}} takes a $s$-domain function \texttt{\textsf{F}} and returns the corresponding time domain model, i.e., the differential equation or its associated solution in the time domain. The other variants that are also used for the transformation of the algebraic expressions in $s$-domain to their corresponding time domain representations are \texttt{\textsf{ilaplace (F,transVar)}} and \texttt{\textsf{ilaplace (F,var,transVar)}}. However, these functions are based on various numerical algorithms that usually involve approximations of the mathematical expressions or results and thus cannot be relied on when performing the analysis of the linear analog circuits used in the safety and mission-critical domains, like transportation, healthcare and energy distribution.

Higher-order-logic theorem proving~\cite{harrison2009handbook}, i.e., a widely used formal verification method, has been used to overcome the inaccuracy limitations of the above-mentioned analysis of the linear analog circuits. It is a computer based mathematical analysis technique that requires developing a mathematical model of the given system in higher-order logic and the formal verification of its intended behaviour, as a mathematically specified property, based on mathematical reasoning within the sound core of a theorem prover. Based on the same motivation, the Laplace transform has been recently formalized in the \holl~theorem prover and it has been utilized to perform the transfer function analysis of the Linear Transfer Converter (LTC) circuit~\cite{taqdees2013formalization}, Sallen-key low-pass filters~\cite{taqdees2017tflac} and a $4$-$\pi$ soft error crosstalk model~\cite{adnan2018JAL}. However, utilizing the Laplace transform theory of \holl~for formally analyzing any system is not a straightforward task as it requires developing a mathematical model and user guidance in the proof process. In this paper, to facilitate engineers for formally analyzing the linear analog circuits, we propose a framework, \fasm, which provides the automatic formal analysis of the Simulink models of these circuits using the formalized Laplace transform theory developed in \holl. The idea is to extract the transfer function of the underlying system after simulating its model developed in Simulink. This extracted transfer function is automatically translated, using a translator developed as a part of this framework, to its corresponding transfer function in \holl, which is then used for the automatic formal analysis of the linear analog circuit using the Laplace transform. Based on our proposed framework, the user only needs to develop the Simulink model of the circuit and the rest of the formal analysis is done almost automatically. For illustration, we use our proposed framework for the automatic formal analysis of the Simulink models of some commonly used analog circuits like Sallen-key filters, etc.


\section{Related Work}\label{SEC:related_work}

To overcome the inherent limitations of the computer-simulation based analysis, formal methods have been previously used for analyzing the Simulink models as well. The MathWorks introduced a verification tool, Simulink Design Verifier (SLDV)~\cite{SDV1}, which is a component of the Simulink environment and is used for formally verifying the Simulink models. SLDV is primarily based on the widely used automated formal verification techniques, i.e., automated theorem proving and model-checking, for the formal verification of the Simulink model of a system and in the case of a failure, it generates a counter-example, which can be used to detect the bug/error in the underlying system's model. Similarly, Silva et al.~\cite{silva2000modeling} provided a computational tool, CheckMate, which is used for formally verifying the hybrid systems. It involves modeling the underlying system using hybrid automata and verification of its associated properties, specified as ACTL formulae, using model checking. Reicherdt et al.~\cite{reicherdt2014formal} proposed to conduct the formal verification of the discrete-time \mlab/Simulink models using Microsoft Boogie program verifier. It involves the automatic translation of \mlab/Simulink models into the Boogie verification language, which allows developing first-order logic based models of the system and their verification using the automated theorem prover Z3. Similarly, Bostrom~\cite{bostrom2011contract} proposed an approach for the contract-based verification of the Simulink models. It involves translating the Simulink models, viewed as Synchronous Data Flow (SDF) graphs, to their corresponding functionally equivalent sequential program statements, which are further analyzed using the automated theorem prover Z3. Joshi et al.~\cite{joshi2005model} proposed an approach for the verification of the discrete time \mlab/Simulink models using the SCADE design verifier. It includes the translation of the \mlab/Simulink models to the synchronous data flow language, Lustre, which are further used for the model-based safety analysis using SCADE. However, due to the inherent computational limitations of the associated formal methods, i.e., less-expressiveness and abstraction in automated theorem proving, and discretization of the continuous models and state-space explosion in model checking, the above-mentioned approaches cannot be termed as accurate and complete when analyzing the complex systems exhibiting continuous dynamics.

Higher-order-logic theorem proving has also been used for formally analyzing the Simulink models. Chen et al.~\cite{chen2009formal} utilized the Prototype Verification System (PVS) theorem prover for the formal verification of Simulink models. Their proposed approach is based on transforming the Simulink models to their corresponding Timed Interval Calculus (TIC) models, which are further analyzed using PVS. However, this approach cannot be utilized for the Laplace transform based analysis of the systems exhibiting the continuous dynamics, which is the main scope of the paper.


\section{Proposed Framework} \label{SEC:prop_framework}

\begin{figure*}[!ht]
\centering
\scalebox{0.585}
{\hspace*{-0.25cm}\includegraphics[trim={5.0 0.0cm 5.0 0.1cm},clip]{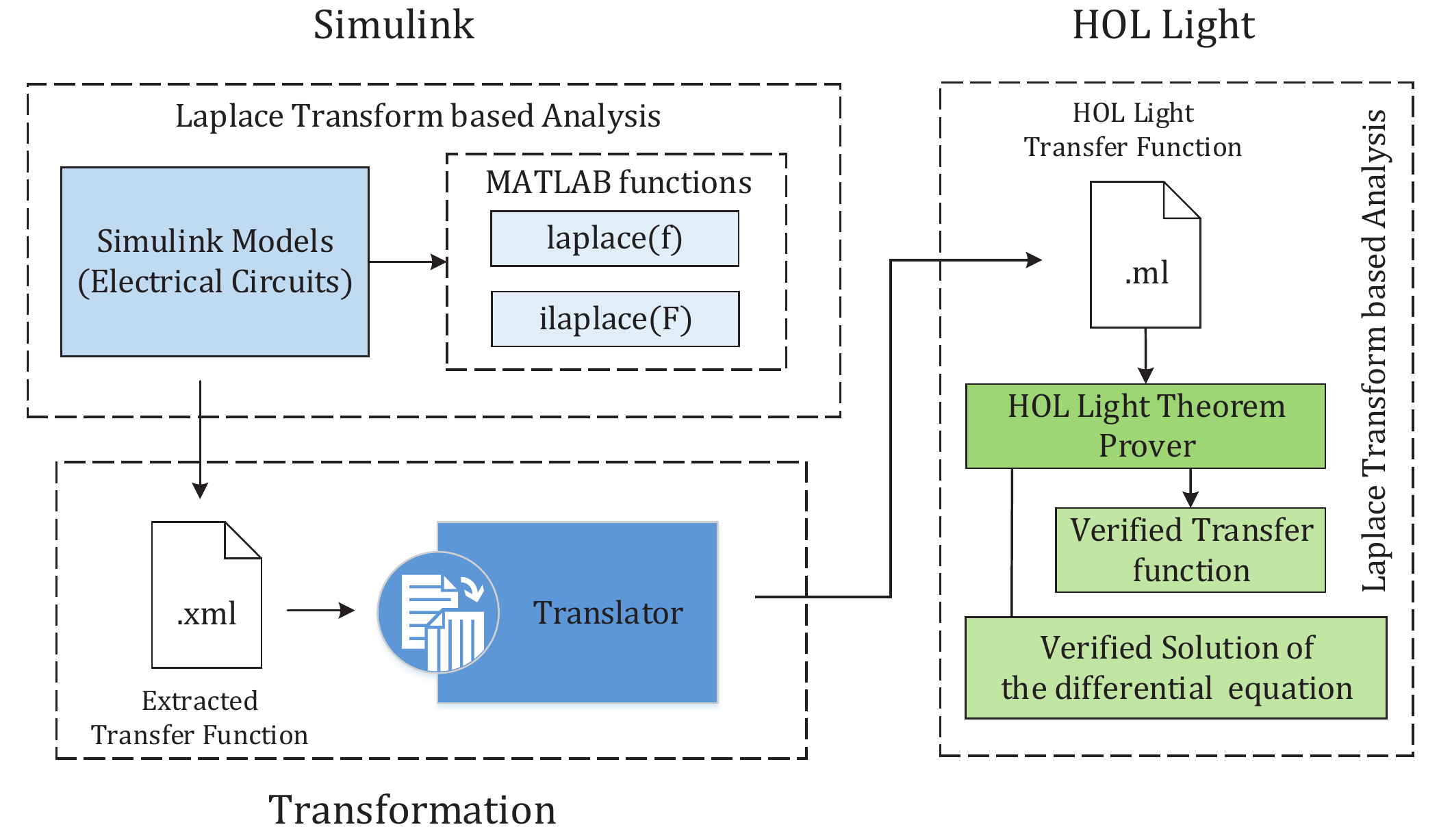}}
\caption{Proposed Framework}
\label{FIG:prop_framework}
\end{figure*}

Our proposed framework, \fasm, for the automatic formal analysis of the Simulink models of the linear analog circuits using the Laplace transform theory of the \holl~theorem prover is depicted in Figure~\ref{FIG:prop_framework}.
The first step is to develop the Simulink model of a linear analog circuit. Next, we need to simulate/analyze the Simulink model to obtain its corresponding transfer function, which represents the input-output relationship in the $s$-domain and is mathematically expressed for the series RLC circuit as:

\begin{equation*}
 \dfrac{V_o(s)}{V_i(s)} = \dfrac{1}{s^2 LC + s RC + 1}
\end{equation*}

Now, in order to perform the formal analysis of a Simulink model, we need to transform this transfer function to its corresponding transfer function in \holl. For this purpose, our framework first extracts the coefficients of the numerator and denominator of the transfer function as lists in \xml. Next, a translator is involved that automatically translates these \xml~lists of coefficients to their corresponding lists in \holl~(\ml). The \holl~lists are further used to formalize the corresponding transfer function and perform the automatic analysis of the linear analog circuits using the Laplace transform theory of \holl~theorem prover, i.e., to formally verify the transfer function and solution of the corresponding time domain models capturing the dynamics of the circuits.
Our proposed framework, \fasm, has twofold advantages: 1) It provides a rigourous analysis of Simulink models of the linear analog circuits unlike their traditional analysis, which is based on the unverified numerical algorithms of \mlab~and thus cannot be trusted for safety-critical applications 2) The translator, developed as a part of this framework, enables the engineers to use the library of the Laplace transform in \holl~for the automatic formal analysis without learning the tool and its associated language, which was not possible in the earlier analysis~\cite{taqdees2013formalization,taqdees2017tflac,adnan2018JAL,rashid2017formal,rashid2017tmformalization,adnan19thesis}.

\section{Preliminaries}\label{SEC:prelim}



\begin{table*}[ht]
	\caption{Laplace Transform}
	\label{TAB:lap_trans_for}
\centering \hspace*{-1.8cm}
  \resizebox{0.850\textwidth}{!}{\begin{minipage}{\textwidth}
{\renewcommand{\arraystretch}{1.005}
	\begin{tabular}{ |p{6.0cm}|p{13.5cm}| }
		\hline 
		Mathematical Form & Formalized Form   \\  \hline \hline
		\multicolumn{2}{|c|}{Laplace Transform} \\ \hline
		{\small {$\begin{array} {lcl} \hspace*{-0.15cm} \mathcal{L} [f(t)] = F(s) =  \\
			\hspace*{1.0cm}	\int_{0}^{\infty} {f(t)e^{-s t}} dt, \ s \ \epsilon \  \mathds{C}
				\end{array}$}  }
		&
		{\small $\begin{array} {lcl} \textup{\texttt{ \hspace*{-0.5cm} $\vdash_{\mathtt{def}}$ $\forall$s f. \kww{laplace\_transform} f s =    }} \\
			\textup{\texttt{   \hspace*{0.0cm}  integral \{t $|$ \&0 $\leq$ drop t\} ($\lambda$t. cexp ($--$(s $\ast$ Cx (drop t))) $\ast$ f t)  }}
			\end{array}$}  \\ \hline
		\multicolumn{2}{|c|}{Laplace Existence} \\ \hline
		\vspace*{-0.5cm}
	{\small	\hspace*{-0.1cm} $f$ is piecewise smooth and is of exponential order on the positive real line}
		&
		{\small$\begin{array} {lcl} \textup{\texttt{  \hspace*{-0.5cm} $\vdash_{\mathtt{def}}$ $\forall$s f. \kww{laplace\_exists} f s $\Leftrightarrow$ }} \\
			\textup{\texttt{ \hspace*{0.0cm} ($\forall$b. f piecewise\_differentiable\_on interval [lift (\&0),lift b]) $\wedge$   }}  \\
			\textup{\texttt{  \hspace*{0.0cm}  ($\exists$M a. Re s $>$ drop a $\wedge$ exp\_order\_cond f M a)   }}
			\end{array}$}  \\ \hline
		\multicolumn{2}{|c|}{Exponential-order Condition} \\ \hline
		\vspace*{-0.45cm}
		{\small \hspace*{-0.1cm} There exist a constant $a$ and a positive constant $M$ such that $|f (t)| \leq Me^{at}$ }
		&
		{ \small $\begin{array} {lcl} \textup{\texttt{  \hspace*{-0.5cm}  $\vdash_{\mathtt{def}}$ $\forall$f M a. \kww{exp\_order\_cond} f M a $\Leftrightarrow$  }} \\
			\textup{\texttt{  \hspace*{0.0cm} \&0 $<$ M $\wedge$  ($\forall$t. \&0 $\leq$ t $\Rightarrow$ norm (f (lift t)) $\leq$ M $\ast$ exp (drop a $\ast$ t))  }}
			\end{array}$}  \\ \hline
		\multicolumn{2}{|c|}{Differential Equation of Order $n$} \\ \hline
		{\small {$\begin{array} {lcl} \hspace*{-0.2cm} \textit{Differential} \ \textit{Equation} = \sum\limits_{k = 0}^{n} {{\alpha}_k \dfrac{d^kf}{{dt}^k}} \end{array}$}  }&
		{\small $\begin{array} {lcl} \textup{\texttt{  \hspace*{-0.5cm}  $\vdash_{\mathtt{def}}$ $\forall$n f t. \kww{diff\_eq\_n\_order} n lst f t =    }} \\
			\textup{\texttt{ \hspace*{0.0cm} vsum (0..n) ($\lambda$k. EL k lst $\ast$  higher\_order\_derivative k f t)  }}
			\end{array}$}  \\ \hline
		\multicolumn{2}{|c|}{Uniqueness of the Laplace Transform} \\ \hline
		\vspace*{-0.9cm}
		If the Laplace transforms of two functions $f$ and $g$ are equal, i.e., $\mathcal{L} [f (t)] = \mathcal{L} [g (t)]$, then both functions $f$ and $g$ are the same, i.e., $f(t) = g(t)$ for all $0 \le t$
		&
	\small	{$\begin{array} {lcl}  \hspace*{-0.35cm} \textup{\texttt{ $\vdash_{\mathtt{thm}}$ $\forall$f g r. \textbf{[A1]} \&0 $<$ Re r $\wedge$   }}  \\
  \textup{\texttt{  \hspace*{-0.35cm}  \textbf{[A2]} ($\forall$s. Re r $\leq$ Re s $\Rightarrow$ laplace\_exists f s) $\wedge$   }}  \\
	\textup{\texttt{ \hspace*{-0.35cm}  \textbf{[A3]} ($\forall$s. Re r $\leq$ Re s $\Rightarrow$ laplace\_exists g s) $\wedge$   }}  \\
	\textup{\texttt{  \hspace*{-0.35cm}  \textbf{[A4]}  ($\forall$s. Re r $\leq$ Re s $\Rightarrow$ laplace\_transform f s = laplace\_transform g s)  }}  \\
			\textup{\texttt{  \hspace*{1.0cm}  $\Rightarrow$  ($\forall$t. \&0 $\leq$ drop t $\Rightarrow$ f t = g t)   }}
			\end{array}$}  \\ \hline
	\end{tabular}
  }
      \end{minipage}}
\end{table*}


This section presents a brief introduction to higher-order-logic theorem proving, the \holl~theorem prover and its Laplace transform theory, which is extensively used in the rest of the paper.

\subsection{Higher-order-logic Theorem Proving}\label{SUBSEC:Theorem_proving}

Theorem proving~\cite{harrison2009handbook} involves constructing the mathematical proofs using a computer program based on axioms and hypothesis. Theorem provers have been extensively used for formalizing classical mathematics, e.g., the formal proof of Kepler conjecture~\cite{hales2005introduction}, the formal library of Euclidean space in \holl~theorem prover~\cite{harrison2013hol} and for verifying many hardware~\cite{camilleri1986hardware} and software~\cite{schumann2001automated} systems. The main idea is to verify a system by formally proving its various properties in the form of mathematical theorems expressed in some appropriate logic, which can be propositional, first-order or higher-order logic.

Based on the decidability or undecidability of the underlying logic, theorem proving can be automatic or interactive. For example, a computer program can automatically verify the theorems about sentences expressed in the propositional logic due to the decidability of this logic whereas the higher-order logic is undecidable and thus verifying proof goals expressed in this logic requires explicit user guidance in an interactive manner.

\subsection{\holl~Theorem Prover}\label{SUBSEC:Hol_light}

\holl~\cite{harrison2009hol} is a higher-order-logic theorem prover that ensures secure theorem proving using the Objective CAML (OCaml) language, which is a variant of the strongly-typed functional programming language \ml.
\holl~has the smallest trusted core (i.e., approximately $400$ lines of OCaml code) compared to all other HOL theorem provers and the underlying logic kernel has been verified in the CakeML project~\cite{kumar2016self}.
\holl~users can interactively verify theorems by applying the available proof tactics and proof procedures. A \holl~theory consist of types, constants, definitions and theorems. \holl~theories are built in a hierarchical fashion and new theories can inherit the definitions and theorems of their parent theories. \holl~consists of a rich set of formalized theories of multivariable calculus and the Laplace transform theories, which are extensively used in the proposed work..
Table~\ref{TAB:lap_trans_for} presents some definitions from the Laplace transform theory of \holl, which includes the Laplace transform, Laplace existence and the exponential-order conditions, the differential equation of order $n$ and the uniqueness of the Laplace transform. Interested readers can refer to~\cite{taqdees2013formalization} for more details about this theory.


\section{\xml~to \holl~(\ml) Translator}\label{SEC:translator}

\begin{figure}[!ht]
\centering
\scalebox{0.420}
{\hspace*{-0.25cm}\includegraphics[trim={5.0 0.2cm 5.0 0.2cm},clip]{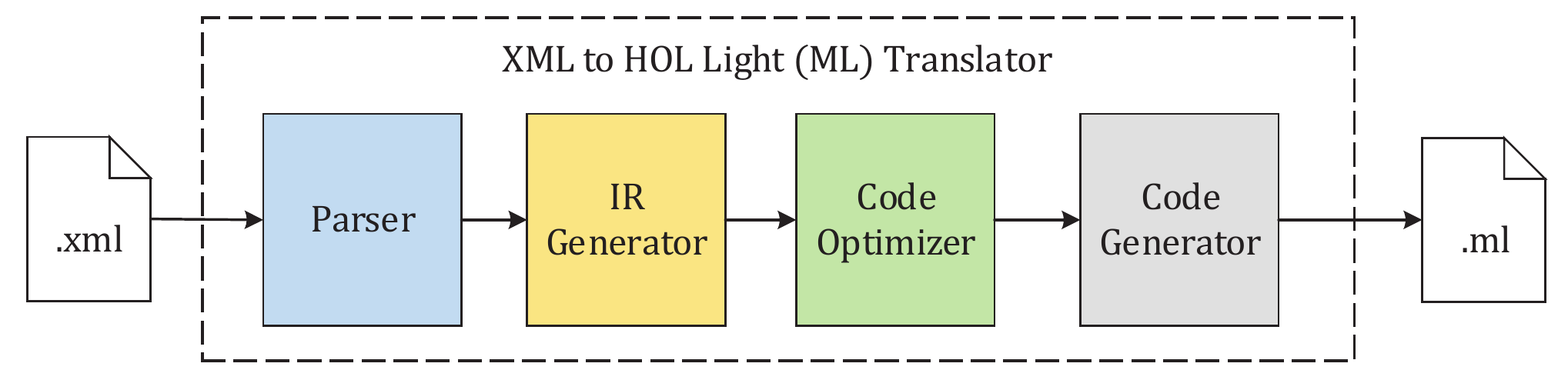}}
\caption{\xml~to \holl~(\ml) Translator}
\label{FIG:xml2mlconverter}
\end{figure}

The main components of our \xml~to \holl~(\ml) translator \footnote{Our translator is available for download for both Windows and Linux operating systems at~\cite{adnan19date}.} are depicted in Figure~\ref{FIG:xml2mlconverter}. It consists of four blocks, namely parser, Intermediate Representation (IR) generator, code optimizer and code generator. Our translator accepts an \xml~file containing the lists of coefficients of the numerator and denominator of the corresponding transfer function of the underlying system. The first block of the translator, i.e., parser, scans the tags and sub-tags of the \xml~file to extract the tags and the contents of the numerator and denominator of the transfer function.

The information extracted from the \xml~file by parser is then provided to the second block, i.e., IR generator, which generates the intermediate representation of the acquired data. This representation involves restructuring of the data in such a way that it facilitates the execution of the subsequent steps towards generating the \ml~representation while retaining the contents of the tags related data.

Next, the intermediate representation produced by the IR generator is passed to the third block, i.e., code optimizer, which optimizes the coefficients of the numerator and denominator. The coefficients extracted by the IR generator have the same syntax as they are represented in \mlab, i.e., scientific (exponential) notation ($E^n$) for the large values of the exponent $n$ (in both command window and workspace) and thus it requires a conversion from the scientific notation to standard form. Moreover, in \holl~(\ml), the decimal numbers and the integer literals of data-type $\mathds{R}$ are represented in two different forms. Thus, in order to make both these representations (\xml~and \holl~representations) coherent, the code optimizer performs these conversions to remodel the IR according to the language of \holl.

Finally, the optimized code is used by the fourth block, i.e., code generator, to generate the output \ml~file, which can be directly used by \holl.

It is to be noted that the purpose of translator is to aid the process of automatic conversion of the Simulink models and it is, itself, not formally verified so it may lead to erroneous translated lists of the coefficients and the associated formal models of the transfer function just like the case with the manual formalization process. However, as the translated model undergoes the formal verification phase using \holl, so there is a high probability of detecting these kinds of flaws during this verification.


\section{Application: Linear Analog Filters}\label{SEC:applications}

Analog filters are used for the purpose of signal filtering and thus allow the passage of signals over a certain range of frequencies. They are of various types, i.e., low-pass, high-pass, band-pass and band-stop etc, based on the range of the frequencies of the input signals that are allowed to pass. For example, a low-pass filter only allows the passage of the signals having frequencies lower than the cut-off frequency and thus reduces the random image noise for a CMOS sensor based camera. To illustrate the practical utilization of our proposed framework, \fasm, we develop a library, which provides the automatic analysis of the Simulink models of various filters, i.e., RC low-pass, first-order all-pass, second-order all-pass, Sallen-key low-pass, Sallen-key high-pass, multiple feedback low-pass, multiple feedback high-pass, multiple feedback band-pass, Boctor notch low-pass and Boctor notch high-pass filters. However, due to the space limitation, we only present the formal analysis of the Simulink model of the Sallen-key low-pass filter.


Sallen-key filters are the linear analog filters, which are extensively used in various applications, such as analog preprocessors, analog-to-digital converters, audio crossover and radio transmitters.
The Simulink model for the Sallen-key low-pass filter is depicted in Figure~\ref{FIG:sec_order_low_pass_filter}. The components $R_1$, $R_2$, $R_3$ and $R_4$ represent the resistors, whereas, $C_1$ and $C_2$ models the corresponding capacitors of the filter. Similarly, $R_L$ represents the load at the output of the filter.
The differential equation modeling the dynamics of the filter for the input voltage $v_i$ and the output voltage $v_o$ is represented as:

\begin{figure*}[!ht]
\centering
\scalebox{0.65}
{\hspace*{-0.25cm}\includegraphics[trim={4.0 0.2cm 1.0 0.2cm},clip]{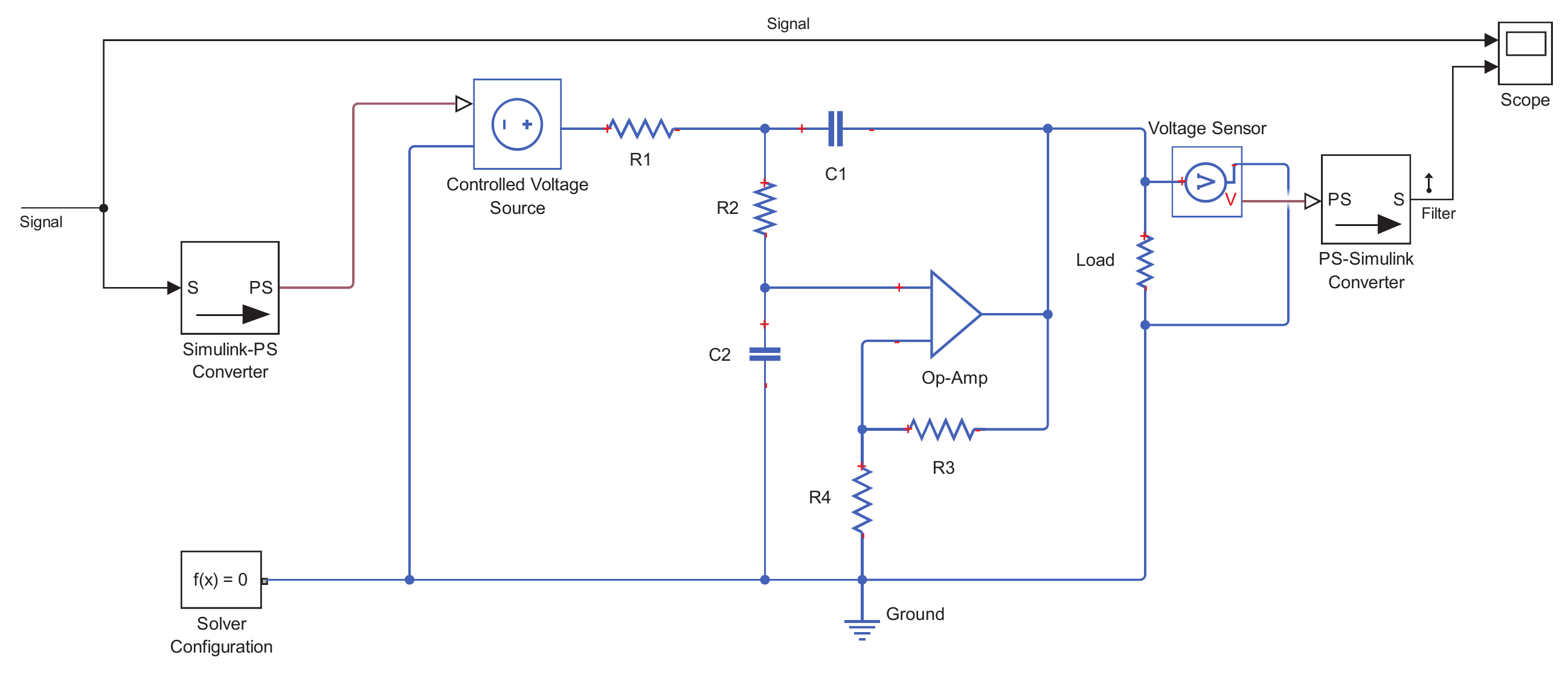}}
\caption{Sallen-key Low-pass Filter}
\label{FIG:sec_order_low_pass_filter}
\end{figure*}


\small

\begin{equation}\label{EQ:diff_eq_low_pass}
\begin{split}
  R_1 R_2 R_4 C_1 C_4 \dfrac{d^2v_o}{dt^2} + (R_1 R_4 C_2 + R_2 R_4 C_2 - R_1 R_3 R_4 C_1) \\
    \dfrac{dv_o}{dt} + R_4 v_o  = (R_3 + R_4) v_i
\end{split}
\end{equation}



\normalsize

Next, we simulate the Simulink model of the filter using the values of its various components as $R_1$ = $R_2$ = $16 K \Omega$, $R_3$ = $R_4$ = $30 K \Omega$, $C_1$ = $C_2$ = $1 n F$, $R_L$ = $10 K \Omega$.
The corresponding transfer function obtained by the simulation of the filter model is mathematically expressed as:



\small

\begin{equation}\label{EQ:transfer_fun_low_pass}
  \dfrac{V_o (s)}{V_i (s)} = \dfrac{7.813 \times {10}^9}{s^2 + 6.25 \times {10}^4 s + 3.906 \times {10}^9}
\end{equation}

\normalsize


\noindent where the coefficients of the numerator and denominator of the transfer function in the above equation are the rounded off values of the coefficients that are obtained from simulations. However, using our proposed framework, we extract their exact values (from \mlab's workspace), i.e., before applying the round-off process, in lists in \xml, i.e., numerator = $[7812500000.48828]$ and denominator = $[1\ 62500.0000039063\ 3906249999.75586]$. These \xml~lists of coefficients are further converted automatically to \holl~lists using our \xml~to \holl~(\ml) translator. Next, these \holl~lists are used to model the corresponding transfer function of the filter. Finally, in order to verify the transfer function, we first model the corresponding differential equation of the filter in \holl~as:


\begin{definition}
\label{DEF:diff_eq_sallen_key_low_pass_filter}
{\footnotesize
\textup{\texttt{\hspace*{-0.15cm} $\vdash_{\mathtt{def}}$ \footnotesize{\kww{inlst\_SKLP\_filter} =  \\
 \hspace*{3.0cm} $\left[\texttt{Cx} \left(\texttt{\#7812500000.48828}\right)\right]$}
}}} \\
{\footnotesize
\textup{\texttt{\hspace*{0.15cm}$\vdash_{\mathtt{def}}$ \footnotesize{\kww{outlst\_SKLP\_filter} = \\
 \hspace*{0.2cm} $\left[\texttt{Cx} \left(\texttt{\#3906249999.75586}\right); \texttt{Cx} \left(\texttt{\#62500.0000039063}\right); \texttt{Cx} \left(\&1\right)\right]$ \vspace*{-0.4cm}  }
}}} \\
{\footnotesize
\textup{\texttt{\footnotesize{\hspace*{0.15cm}$\vdash_{\mathtt{def}}$ \kww{diff\_eq\_SKLP\_filter} VI V0 $\Leftrightarrow$ \\
$\mathtt{\ }$\hspace{-0.2cm} ($\forall$t. diff\_eq\_n\_order 2 outlst\_SKLP\_filter V0 t = \\
$\mathtt{\ }$\hspace{0.65cm} diff\_eq\_n\_order 0 inlst\_SKLP\_filter VI t) }}}}
\end{definition}

%

\noindent where the \holl~function \texttt{diff\_eq\_n\_order} models the linear differential equation of order $n$, given in Table~\ref{TAB:lap_trans_for}. The symbol \texttt{\#} is used to represent a decimal number of data-type $\mathds{R}$ in \holl.

Now, we formally verify the corresponding transfer function as the following \holl~theorem:


\begin{theorem}
\label{THM:transfer_fun_sklp_filter}
{\footnotesize
\textup{\texttt{$\vdash_{\mathtt{thm}}$ $\forall$VI V0 s.   \\
	$\mathtt{\ }$\hspace{-0.1cm} \textbf{[A1]}  ($\forall$t. differentiable\_higher\_deriv 0 VI t) $\wedge$   \\
	$\mathtt{\ }$\hspace{-0.1cm} \textbf{[A2]}  ($\forall$t. differentiable\_higher\_deriv 2 V0 t) $\wedge$   \\
	$\mathtt{\ }$\hspace{-0.1cm} \textbf{[A3]} zero\_init\_conditions 1 V0 $\wedge$  \\
	$\mathtt{\ }$\hspace{-0.1cm} \textbf{[A4]} laplace\_transform VI s $\neq$ Cx (\&0) $\wedge$  \\
	$\mathtt{\ }$\hspace{-0.1cm} \textbf{[A5]} \big($\texttt{s}^{\texttt{2}}$ \texttt{+ Cx} $\left(\texttt{\#62500.0000039063}\right)$ $\ast$ \texttt{s +}  \\
   \hspace*{2.0cm}  \texttt{Cx} $\left(\texttt{\#3906249999.75586}\right)$ $\neq$ Cx (\&0)\big) $\wedge$  \\
	$\mathtt{\ }$\hspace{-0.1cm} \textbf{[A6]} laplace\_exists\_higher\_deriv 0 VI s $\wedge$   \\
	$\mathtt{\ }$\hspace{-0.1cm} \textbf{[A7]} laplace\_exists\_higher\_deriv 2 V0 s $\wedge$   \\
	$\mathtt{\ }$\hspace{-0.1cm} \textbf{[A8]} ($\forall$t. diff\_eq\_SKLP\_filter VI V0 t) $\wedge$  \\
 $\mathtt{\ }$\hspace{1.0cm} \vspace{-0.2cm} \\
	$\mathtt{\ }$\hspace{1.0cm} $\Rightarrow$ $\dfrac{\texttt{laplace\_transform V0 s}}{\texttt{laplace\_transform VI s}}$ = \\
 $\mathtt{\ }$\hspace{1.0cm} \vspace*{0.0cm} \\
	$\mathtt{\ }$\hspace{-0.2cm} \footnotesize{$\dfrac{\texttt{Cx} \left(\texttt{\#7812500000.48828}\right)}{\texttt{s}^{\texttt{2}} \texttt{+ Cx} \left(\texttt{\#62500.0000039063}\right) \ast \texttt{s + Cx} \left(\texttt{\#3906249999.75586}\right)  }$}
}}}
\end{theorem}

The assumptions \texttt{A1-A2} provide the differentiability conditions for the input \texttt{VI} and output \texttt{V0} up to the order $0$ and $2$, respectively. The assumption \texttt{A3} presents the \textit{zero initial conditions} for the function \texttt{V0}. The assumptions \texttt{A4-A5} express the design constraints for the Sallen-key low-pass filter. Similarly, the assumptions \texttt{A6-A7} ensure that the Laplace transform of the functions \texttt{VI} and \texttt{V0} exist up to order $0$ and $2$, respectively. The assumption \texttt{A8} provides the differential equation model of the underlying filter.  Finally, the conclusion represents the corresponding transfer function (Equation~(\ref{EQ:transfer_fun_low_pass})). The verification of Theorem~\ref{THM:transfer_fun_sklp_filter} is done automatically using the automatic tactic \texttt{DIFF\_EQ\_2\_TRANS\_FUN\_TAC}, which is developed as a part of our theory of the Laplace transform in~\holl.
It requires the differential equation and the transfer function of the underlying system and automatically verifies the theorem corresponding to the transfer function of the system.

Next, we formally verify the differential equation of the Sallen-key low-pass filter based on its transfer function as:

\begin{theorem}
\label{THM:diff_eq_sklp_filter}
{
\footnotesize
\textup{\texttt{$\vdash_{\mathtt{thm}}$ $\forall$VI V0 r.  \\
	$\mathtt{\ }$\vspace*{0.00cm}\hspace*{-0.3cm} \textbf{[A1]}  ($\forall$t. differentiable\_higher\_deriv 0 VI t) $\wedge$   \\
	$\mathtt{\ }$\vspace*{0.00cm}\hspace*{-0.3cm} \textbf{[A2]}  ($\forall$t. differentiable\_higher\_deriv 2 V0 t) $\wedge$ \\
	$\mathtt{\ }$\vspace*{0.00cm}\hspace*{-0.3cm}  \textbf{[A3]} zero\_init\_conditions 1 V0 $\wedge$  \\
	$\mathtt{\ }$\vspace*{0.00cm}\hspace*{-0.3cm} \textbf{[A4]} \big($\forall$s. Re r $\leq$ Re s $\Rightarrow$ \big($\texttt{s}^{\texttt{2}}$ \texttt{+ Cx} $\left(\texttt{\#62500.0000039063}\right)$   \\
   \hspace*{1.2cm}  $\ast$ \texttt{s + } \texttt{Cx} $\left(\texttt{\#3906249999.75586}\right)$ $\neq$ Cx (\&0)\big)\big) $\wedge$  \\
	$\mathtt{\ }$\vspace*{0.00cm}\hspace*{-0.3cm} \textbf{[A5]} ($\forall$s. Re r $\leq$ Re s $\Rightarrow$   \\
  \hspace*{2.0cm}   laplace\_transform VI s $\neq$ Cx (\&0)) $\wedge$ \\
	$\mathtt{\ }$\vspace*{0.00cm}\hspace*{-0.3cm} \textbf{[A6]} \&0 $<$ Re r $\wedge$  \\
	$\mathtt{\ }$\vspace*{0.00cm}\hspace*{-0.3cm} \textbf{[A7]} ($\forall$s. Re r $\leq$ Re s $\Rightarrow$  \\
  \hspace*{2.0cm}   laplace\_exists\_higher\_deriv 0 VI s) $\wedge$   \\
	$\mathtt{\ }$\vspace*{0.0cm}\hspace{-0.3cm} \textbf{[A8]} ($\forall$s. Re r $\leq$ Re s $\Rightarrow$  \\
 \hspace*{2.0cm}  laplace\_exists\_higher\_deriv 2 V0 s) $\wedge$ \vspace*{0.00cm}  \\
	$\mathtt{\ }$\hspace{-0.3cm} \textbf{[A9]} ($\forall$s. Re r $\leq$ Re s $\Rightarrow$ $\dfrac{\texttt{laplace\_transform V0 s}}{\texttt{laplace\_transform VI s}}$ = \\  \\
	$\mathtt{\ }$\vspace*{0.00cm}\hspace*{-0.35cm} $\dfrac{\texttt{Cx} \left(\texttt{\#7812500000.48828}\right)}{\texttt{s}^{\texttt{2}}\ \texttt{+ Cx} \left(\texttt{\#62500.0000039063}\right) \ast \texttt{s + Cx} \left(\texttt{\#3906249999.75586}\right)  }$\Bigg) \vspace*{0.0cm} \\
	$\mathtt{\ }$\vspace*{0.00cm}\hspace*{0.0cm} $\Rightarrow$ ($\forall$t. \&0 $\leq$ drop t. diff\_eq\_SKLP\_filter VI V0 t)
}}}
\end{theorem}

The assumptions \texttt{A1-A5} are the same as that of Theorem~\ref{THM:transfer_fun_sklp_filter}. The assumption \texttt{A6} ensures that the real part of the Laplace variable \texttt{r} is always positive.
The assumptions \texttt{A7-A8} ensure that the Laplace transform of the functions \texttt{VI} and \texttt{V0} exist up to order $0$ and $2$, respectively. The assumption \texttt{A9} provides the transfer function of the Sallen-key low-pass filter. Finally, the conclusion provides its corresponding differential equation model.
The verification of Theorem~\ref{THM:diff_eq_sklp_filter} is done automatically using the automatic tactic \texttt{TRANS\_FUN\_2\_DIFF\_EQ\_TAC}, which is developed as a part of our theory of the Laplace transform in \holl.
It requires the differential equation and the transfer function of the underlying system and automatically verifies the theorem corresponding to the differential equation of the system.

The details about the formal analysis of the Simulink models of the Sallen-key filters, and the other analog filters can be found at~\cite{adnan19date}. The distinguishing feature of the automatic formal analysis of the analog filters is that all the assumptions are explicitly present in the analysis as missing any necessary condition would not allow the verification of the theorem due to the soundness of \holl~theorem prover.
On the other hand, the traditional analysis techniques used in Simulink cannot provide this guarantee due to their informal nature of analysis. It is important to note that not catering for any such assumption during the design time may lead to an erroneous design and thus to some undesired consequences. Moreover, our automatic tactics \texttt{DIFF\_EQ\_2\_TRANS\_FUN\_TAC} and \texttt{TRANS\_FUN\_2\_DIFF\_EQ\_TAC} are of generic nature and can be used for the automatic verification of a linear analog filter of any order. Thus, there is no restriction on the size and order of the linear analog circuit.
This way \fasm~provides all the useful features of a theorem proving based analysis in an automatic manner.


\section{Conclusion}\label{SEC:conclusion}

This paper presents a Framework for the Automatic formal analysis of Simulink Models (\fasm) of the linear analog circuits, which are widely used in signal processing architectures. \fasm~provides rigorous analysis of analog circuits due to the involvement of a higher-order-logic theorem prover. As a part of this framework, we developed a translator, which automatically translates the transfer function obtained from simulating the Simulink model of a given linear analog circuit to its corresponding \holl~representation. Moreover, this transfer function is automatically verified using our Laplace transform theory of the \holl~theorem prover. Finally, we use \fasm~to perform the automatic formal analysis of Simulink models of the commonly used analog filters. It is important to note that \fasm~can equally be used for formally analyzing any non-linear analog circuit as long as the non-linear circuit can be linearized first~\cite{ochotta2012practical}.


\bibliographystyle{IEEEtran}
\bibliography{biblio}
%
%
\end{document}